\documentclass{article}
\usepackage{epsfig}

\newread\epsffilein    
\newif\ifepsffileok    
\newif\ifepsfbbfound   
\newif\ifepsfverbose   
\newdimen\epsfxsize    
\newdimen\epsfysize    
\newdimen\epsftsize    
\newdimen\epsfrsize    
\newdimen\epsftmp      
\newdimen\pspoints     
\def\epsfbox#1{\global\def\epsfllx{72}\global\def\epsflly{72}%
   \global\def\epsfurx{540}\global\def\epsfury{720}%
   \def\lbracket{[}\def\testit{#1}\ifx\testit\lbracket
   \let\next=\epsfgetlitbb\else\let\next=\epsfnormal\fi\next{#1}}%
\def\epsfgetlitbb#1#2 #3 #4 #5]#6{\epsfgrab #2 #3 #4 #5 .\\%
   \epsfsetgraph{#6}}%
\def\epsfnormal#1{\epsfgetbb{#1}\epsfsetgraph{#1}}%
\def\epsfgetbb#1{%

\openin\epsffilein=#1
\ifeof\epsffilein\errmessage{I couldn't open #1, will ignore it}\else

   {\epsffileoktrue \chardef\other=12
    \def\do##1{\catcode`##1=\other}\dospecials \catcode`\ =10
    \loop
       \read\epsffilein to \epsffileline
       \ifeof\epsffilein\epsffileokfalse\else

          \expandafter\epsfaux\epsffileline:. \\%
       \fi
   \ifepsffileok\repeat
   \ifepsfbbfound\else
    \ifepsfverbose\message{No bounding box comment in #1; using defaults}\fi\fi
   }\closein\epsffilein\fi}%

\def\epsfsetgraph#1{%
   \epsfrsize=\epsfury\pspoints
   \advance\epsfrsize by-\epsflly\pspoints
   \epsftsize=\epsfurx\pspoints
   \advance\epsftsize by-\epsfllx\pspoints

   \epsfsize\epsftsize\epsfrsize
   \ifnum\epsfxsize=0 \ifnum\epsfysize=0
      \epsfxsize=\epsftsize \epsfysize=\epsfrsize

     \else\epsftmp=\epsftsize \divide\epsftmp\epsfrsize
       \epsfxsize=\epsfysize \multiply\epsfxsize\epsftmp
       \multiply\epsftmp\epsfrsize \advance\epsftsize-\epsftmp
       \epsftmp=\epsfysize
       \loop \advance\epsftsize\epsftsize \divide\epsftmp 2
       \ifnum\epsftmp>0
          \ifnum\epsftsize<\epsfrsize\else
             \advance\epsftsize-\epsfrsize \advance\epsfxsize\epsftmp \fi
       \repeat
     \fi
   \else\epsftmp=\epsfrsize \divide\epsftmp\epsftsize
     \epsfysize=\epsfxsize \multiply\epsfysize\epsftmp
     \multiply\epsftmp\epsftsize \advance\epsfrsize-\epsftmp
     \epsftmp=\epsfxsize
     \loop \advance\epsfrsize\epsfrsize \divide\epsftmp 2
     \ifnum\epsftmp>0
        \ifnum\epsfrsize<\epsftsize\else
           \advance\epsfrsize-\epsftsize \advance\epsfysize\epsftmp \fi
     \repeat
   \fi

   \ifepsfverbose\message{#1: width=\the\epsfxsize, height=\the\epsfysize}\fi
   \epsftmp=10\epsfxsize \divide\epsftmp\pspoints
   \vbox to\epsfysize{\vfil\hbox to\epsfxsize{%

      \special{illustration #1 scaled \number\epsfscale}
      \hfil}}%
\epsfxsize=0pt\epsfysize=0pt\epsfscale=1000 }%

{\catcode`\%=12 \global\let\epsfpercent=

\long\def\epsfaux#1#2:#3\\{\ifx#1\epsfpercent
   \def\testit{#2}\ifx\testit\epsfbblit
      \epsfgrab #3 . . . \\%
      \epsffileokfalse
      \global\epsfbbfoundtrue
   \fi\else\ifx#1\par\else\epsffileokfalse\fi\fi}%

\def\epsfgrab #1 #2 #3 #4 #5\\{%
   \global\def\epsfllx{#1}\ifx\epsfllx\empty
      \epsfgrab #2 #3 #4 #5 .\\\else
   \global\def\epsflly{#2}%
   \global\def\epsfurx{#3}\global\def\epsfury{#4}\fi}%

\newcount\epsfscale    
\newdimen\epsftmpp     
\newdimen\epsftmppp    
\newdimen\epsfM        
\newdimen\sppoints     
\epsfscale=1000        
\sppoints=1000sp       
\epsfM=1000\sppoints

\def\computescale#1#2{%
  \epsftmpp=#1 \epsftmppp=#2
  \epsftmp=\epsftmpp \divide\epsftmp\epsftmppp  
  \epsfscale=\epsfM \multiply\epsfscale\epsftmp 
  \multiply\epsftmp\epsftmppp                   
  \advance\epsftmpp-\epsftmp                    
  \epsftmp=\epsfM                               
  \loop \advance\epsftmpp\epsftmpp              
    \divide\epsftmp 2                           
    \ifnum\epsftmp>0
      \ifnum\epsftmpp<\epsftmppp\else           
        \advance\epsftmpp-\epsftmppp            
        \advance\epsfscale\epsftmp \fi          
  \repeat
  \divide\epsfscale\sppoints}
\def\epsfsize#1#2{%
  \ifnum\epsfscale=1000
    \ifnum\epsfxsize=0
      \ifnum\epsfysize=0
      \else \computescale{\epsfysize}{#2}
      \fi
    \else \computescale{\epsfxsize}{#1}
    \fi
  \else
    \epsfxsize=#1
    \divide\epsfxsize by 1000 \multiply\epsfxsize by \epsfscale
  \fi}

\font\tenbg=cmmib10 at 10pt

\def \rvecphi{{\hbox{\tenbg\char'036}}}

\begin{document}
\title{Jets and Disk-Winds from Pulsar Magnetospheres}

\author{
R.V.E. Lovelace$^1$, L. Turner$^2$,
and  M.M. Romanova$^3$}

\maketitle

\begin{abstract}

We discuss axisymmetric
force-free pulsar magnetospheres with
magnetically collimated jets and a disk-wind
obtained by numerical solution of the pulsar
equation.
  This solution represents an alternative
to the quasi-spherical wind solutions where a
major part of the
current flow is  in a current sheet which
is unstable to magnetic field annihilation.

\noindent(1) {Departments of Astronomy and  Applied and
Engineering Physics,
Cornell University, Ithaca, NY 14853-6801;
RVL1@cornell.edu}

\noindent(2) {Department of Astronomy,
Cornell University, Ithaca, NY 14853-6801; lt79@cornell.edu}

\noindent(3) {Department of Astronomy,
Cornell University, Ithaca, NY 14853-6801; romanova@astro.cornell.edu}

\end{abstract}

\noindent{Subject Headings: stars: neutron --- pulsars: general
--- stars: magnetic fields --- X-rays: stars}

\section{Introduction}

    Interest in the structure of pulsar magnetospheres
has been stimulated by Chandra and Hubble Space
Telescope observations of the Crab synchrotron nebula
which point to an axial-jet/equatorial-disk structure
(Hester et al. 2002).
   An analogous structure is
observed in the nebula of the Vela pulsar
(Pavlov et al. 2001).
  A theoretical model of an aligned rotating pulsar
with collimated jets was put forward by
Sulkanen and Lovelace (1990; SL) who solved the
pulsar equation of Scharlemann and Wagoner (1973) on a grid
numerically.
  This work was criticized by Contopoulos, Kazanas,
and Fendt (1999;  CKF) who presented numerical calculations
showing that the
possibly unique solution is a
quasi-spherical wind {\it without} jets but
with an equatorial current sheet.
   The quasi-spherical wind  solution has been found in
the time-dependent, relativistic, force-free simulations by
Spitkovsky (2004), Komissarov (2006), and McKinney (2006)
and  in high resolution grid calculations
by Timokhin (2006).
  However, the wind solutions may be short-lived
owing to the fact that a major part of the current
flow is in a current sheet which is unstable to magnetic
field annihilation.
MHD simulations by Komissarov and
Lyubarsky (2004, 2006) indicate that a jet-torus configuration
can be generated due to the anisotropic
energy flux density of the pulsar far outside the light cylinder.

    We reconsider the possibility of  jet/disk-wind
structures of aligned pulsar magnetospheres on the
scale of the light-cylinder distance
using a different approach to the numerical solution
of the pulsar equation.
   We utilize the fact that the poloidal current
flow along the poloidal field lines within the
star's light-cylinder is an adjustable parameter.
  We find that when this parameter is sufficiently
large,  magnetically collimated ($\pm z$) jets form  within
the light-cylinder and a quasi-collimated flows outside.
  The collimation is due to the toroidal magnetic field.
   The analysis by SL did not include the
current flows outside the light-cylinder and
this resulted in a kink in the field lines which cross
the light-cylinder.
  In addition to the
collimated  flows, we find an
``anti-collimated'' disk-wind.
  The anti-collimation is due to
the toroidal magnetic field.
   These solutions have no net poloidal current
flow and no current sheets inside or outside the
light-cylinder.
   Thus these solutions are not unstable to field
annihilation.
   The formation of magnetically collimated
jets along the axes and an equatorial disk-wind
is similar to what is found in the nonrelativistic
limit for magnetic loops
threading an accretion disk (Ustyugova et al. 2000).
   This jet/disk-wind geometry was discussed for the case
of pulsars by Romanova, Chulsky, and Lovelace (2005).

    Section 2 of the paper discusses the theory, the boundary
conditions, and the regularity condition at the light-cylinder.
It goes on to discuss the conditions for having no
jets and having jets.    Section 3 discusses
the numerical solutions.
    For the case of jets we discuss the radial force
balance across the jet and the vertical force balance
in the disk.
   Section 4 gives the conclusions of this work.

\section{Theory}

   The main equations for the plasma follow from the
continuity equation ${\nabla \cdot}(\rho{\bf v})=0$,
Amp\`ere's law, ${\bf \nabla \times B}=4\pi {\bf J}/c$,
Coulomb's law ${\bf \nabla \cdot E}= 4\pi \rho_e$,
with $\rho_e$ the charge density, Faraday's law,
${\bf \nabla \times E}=0$, perfect conductivity,
${\bf E}+{\bf v \times B}/c=0$, with ${\bf v}$ the
plasma flow velocity, and the ``force-free'' condition
in the Euler equation, $\rho_e {\bf E}
+{\bf J \times B}/c=0$.
  The perfect conductivity implies
that ${\bf E}^2 <{\bf B}^2$.
   Owing to the assumed axisymmetry, $E_\phi=0$, so
that the poloidal velocity ${\bf v}_p =
\kappa {\bf B}_p$.
   Mass conservation then gives
${\bf B}\cdot {\bf \nabla} (\rho \kappa) =0,$
which implies that $\rho \kappa = F(\Psi)/4\pi$,
where $F$ is an arbitrary function of the flux function $\Psi$.
 In cylindrical coordinates,
$B_r=-r^{-1}(\partial \Psi/\partial z)$
and $B_z=r^{-1}(\partial \Psi/\partial r)$.
   In a similar way one finds that
$v_\phi -\kappa B_\phi =r G(\Psi)$, so
that ${\bf E} = - G(\Psi) {\bf \nabla} \Psi$,
and $rB_\phi = H(\Psi)$, so that there are  two
additional functions, $G$ and $H$.

    The function $G$ is determined along
all of the field lines which go through
the star.
   This follows
from the perfect conductivity condition
at the surface of the star,
    $E_\theta +({\bf v \times B})_\theta/c=0$,
in terms of spherical $(R,\theta,\phi)$
coordinates.
   This gives
$E_\theta=-(v_\phi B_R - v_R B_\phi)/c =
-v_\phi B_R/c$, where $v_R$ is
zero inside the star.
  Here, $v_\phi=\Omega_*R_*\sin\theta$ is the
velocity at the star's surface, $\Omega_*$
is the angular velocity of the star, and
$R_*$ is the star's radius.
  Thus we have
$E_\theta(R_*,\theta)= -\Omega_*~ (d\Psi/d\theta)/(R_*c)$,
so that $G(\Psi)=\Omega_*/c$.

  The component of the Euler equation in
the direction of ${\bf \nabla }\Psi$ gives
the force-free
Grad-Shafranov (GS) equation in
cylindrical $(r,\phi,z)$ coordinates,
\begin{equation}
\left[1-\left({r\Omega_* \over c}
\right)^2\right]\Delta^\star \Psi
-{2 r \Omega_*^2 \over c^2}{\partial \Psi \over \partial r}
=-{\cal F}(\Psi)~,
\end{equation}
where
 $$
{\cal F} \equiv H(\Psi) {dH(\Psi)\over d\Psi}~,
$$
and
$$
 \Delta^\star \equiv
{\partial^2 / \partial r^2}
-(1/r)(\partial / \partial r)
+{\partial^2 / \partial z^2}~,
$$
(Scharlemann \& Wagoner 1973).

    Note that
the poloidal current density is
given as ${\bf J}_p =(c/4\pi){\bf B}_p(dH/d\Psi)$
where the $p-$subscript indicates
the poloidal component.
   We consider solutions with symmetry about
the equatorial plane with for example
$\Psi(r,z)=\Psi(r,-z)$ and $B_\phi(r,z)=-B_\phi(r,-z)$.

   This equation for $\Psi$ involves
the unknown function ${\cal F}(\Psi)$
or $H(\Psi)$.
 It is in general
nonlinear.
   Amp\`ere's law gives
$\oint d{\bf l}\cdot {\bf B}
=(4\pi/c)\int d{\bf S}\cdot {\bf J}$,
so that $rB_\phi(r,z)=H(\Psi)$ is $(2/c)$ times
the current flowing through a circular
area of radius $r$ (with normal $\hat{\bf z}$)
labeled by $\Psi(r,z)$= const.

 In the following distances are measured in units
of the light cylinder radius, $r_L = c/\Omega_*$.
The flux function $\Psi$ is measured in units
of $\mu/r_L$, where $\mu$ is the
magnetic moment of the star.
   The magnetic field is measured in units
of $B_0\equiv \mu/r_L^3$.

\subsection{Boundary Conditions}

Numerical solutions of equation (1) for $\Psi(r,z)$
are calculated on
a uniform grid in a  region $r=0$ to $r_{\rm max}$
and $z=0$ to $z_{\rm max}$ with $r_{\rm max} = z_{\rm max} \gg 1$.
For $r^2+z^2 \ll 1,$ we require
$\Psi \rightarrow (r^2+z^2)^{-1/2}$
which is the star's intrinsic dipole field.
Along the symmetry axis, $\Psi(r=0,z)=0$,
and $H(\Psi=0) = 0$.

   On the equatorial plane inside the light
cylinder ($r<1,~z=0$),
$B_r=-(1/r)(\partial \Psi/\partial z)=0$
and $H(r,z)=0$.  For the closed field lines
within the light cylinder $\Psi > \Psi_{eq}$.
   On the equatorial plane outside the light
cylinder ($r>1,~z=0$), $\Psi(r,0)=\Psi_{eq}=$ const
and $H(r,0)=0$.  Thus for the open field lines, the
range of $\Psi$ is from zero on the axis to
$\Psi_{eq}$ on the equatorial plane.

   On the outer boundaries at $r_{\rm max}$ and
$z_{\rm max}$, we take free boundary conditions
$\partial^2\Psi/\partial n^2 =0$ where $n$ is
the normal to the boundary.
   Other conditions have been tested including
using equation (1) and they do not alter
our results.

\subsection{Light Cylinder Condition}

   A further condition on the solutions of equation
(1) arises from the regularity of $\Psi$ at the
light cylinder.
Because the coefficient of the $\Delta^\star$ term vanishes
at the light cylinder,
 all field lines ($\Psi$ values) which
cross the light cylinder must
have
\begin{equation}
2{\partial \Psi \over \partial r} =
H{d H \over d \Psi}={\cal F}\quad {\rm for}~~
r=1~{\rm and}~ {\rm all}~ z~.
\end{equation}
This relation determines $H(\Psi)$
for the field lines which cross the light-cylinder.
In the following $dH/d\Psi =H^\prime$.

There are  two possibilities:

\subsubsection{No Jets}

 One is that all open field lines cross
the light-cylinder
This is the solution put forward by CKF.

   Because we calculate $\Psi$ in a finite size
region, ${\cal F}(\Psi)$  is not
determined for the open field lines
which exit the region at $z=z_{\rm max}$
inside the light-cylinder,
$r=0$ to $1$.
  These field lines have $0\leq \Psi < \Psi_c$,
where $\Psi_c \equiv \Psi(1,z_{\rm max})$.
 For the CFK solution, for  this range of $\Psi$
we assume a linear
interpolation, ${\cal F}=
(\Psi/\Psi_c){\cal F}_c$, where the $c$-subscript
indicates evaluation at $(r=1,~z=z_{\rm max})$.
  The quantity ${\cal F}_c$ is known owing to
equation (2).  Note that both $\Psi_c$ and
${\cal F}_c$ evolve as the iteration proceeds.

   For the CKF case,
\begin{equation}
\int_0^{\Psi_{eq}-0^+} d\Psi~ {\cal F}=
{1\over 2}H^2(\Psi_{eq}-0^+)>0~,
\end{equation}
where $0^+$ is an arbitrarily small positive quantity.
Because $H(\Psi_{eq})=0$, this requires
a poloidal current
sheet  at $z=0^+$ for $r>1$ with
\begin{eqnarray}
J_r &=&-{c\over 4\pi} B_r(r, 0^+) H(\Psi_{eq}-0^+)
\delta(\Psi-\Psi_{eq}+0^+)~,
\nonumber \\
&=&-{c \over 4\pi r}H(\Psi_{eq}-0^+)\delta(z-0^+)~.
\end{eqnarray}
The poloidal current flow is sketched in
Figure 1a.
   There is a corresponding positive current density
$\propto \delta(z+0^+)$ due to the lower half-space.
  Inside the light-cylinder
the poloidal current
sheet follows the dipole-like poloidal field
lines to the star's surface.  This is required
in order to have $H=0$ within the closed field
line region of the magnetosphere.  At the
same time there is a delta-function
toroidal current flow with
\begin{equation}
J_\phi =(c/4\pi)B_r(r,0^+)\delta(z-0^+)
\end{equation}
for $r>1$.
   There is a corresponding positive toroidal
current flow $\propto \delta(z+0^+)$ due to
the lower half-space.
  Inside
the light-cylinder there is also a delta-function
toroidal current sheet associated with the
mentioned poloidal current sheet.

  The magnetic field just above the
equatorial plane for $r>1$ is ${\bf B}_+=B_r \hat{\bf {r}}+
B_\phi \hat{\rvecphi~}$, and this is equal and opposite
to the field just below the plane.
  The oppositely directed fields are unstable
to annihilation.
   The electromagnetic stress $T_{zz}$ varies
discontinuously from zero  on
the equatorial plane (by symmetry) to
a finite value at $z=0^+$.

\subsubsection{Jets}

   A second possibility is that
there is a collimated jet along the
$+z$ axis (and $-z$ axis) so
that not all open field lines
cross the light cylinder.
  The open field lines
which do cross the light cylinder must
obey equation (2).
For the other open but
collimated field lines
($0\leq r \leq r_L$,~ $z \rightarrow \infty$),  $H(\Psi)$
is arbitrary.

    Thus $H(\Psi)$ is
 not determined for
$\varphi \equiv \Psi/\Psi_c  <1$.
Because $H(0)=0$ we  consider the simple dependence
\begin{equation}
H(\Psi)=k_H\left(\varphi -
{1\over 2}\beta \varphi^2\right)~,
\end{equation}
where $k_H$ is an adjustable
constant.
   As $k_H$ increases in equation (6)
the  axial current flow  within the
light-cylinder ($\propto dH/d\Psi$) increases.
   Owing to equation (2),
$(HH^\prime)_c$
is fixed at each step of
the numerical iteration of $\Psi$.
Of course, both $\Psi_c$
and $(HH^\prime)_c$ evolve as the
iteration proceeds.
Consequently, we can calculate
$$
\beta = {1\over 2}\left[3-\left(1+{8\Psi_c(HH^\prime)_c
\over k_H^2}\right)^{1/2}\right]~,
$$
at each iteration.

   We seek solutions {\it without} an
equatorial current sheet.  That is,
we search for solutions with
\begin{equation}
\int_0^{\Psi_{eq}} d\Psi~{\cal F} =0~.
\end{equation}
Figure 1b shows a sketch of the
poloidal current flow.

\section{Numerical Solutions}

   The numerical calculations of $\Psi$ for
both cases of no jets and jets were done using
successive iterations
on a uniform $500 \times 500$ mesh in
a square region $r_{\rm max}\times z_{\rm max}
=5\times 5$.
   We have obtained the same results
for $600\times 600$ grid for a $6\times 6$ region.
   The initial  $\Psi(r,z)$ used to start
the iteration consists of the vacuum dipole
field inside the light-cylinder and the straight
line extrapolation of the field lines outside
this cylinder.
Convergence of the iterations is measured by the change
of $\Psi$ between iterations.
Figure 2 shows the solution $\Psi(r,z)$
and $H(\Psi)$ for the
CKF case where there is no jet.

\subsection{Solutions with Jets}

    Figure 3 shows $\Psi(r,z)$ and $H(\Psi)$
for the case of a jet flow along the axis and
a disk-wind in the equatorial plane where
$k_H=2.48$.
The flow has
 zero net poloidal current flow,
$\int_0^{\Psi_{eq}} d\Psi {\cal F}=0$ and
{\it no}  current sheets.
   The main parameters of this solution are:
along the $z-$axis $\Psi=0=H$; on the light-cylinder
(at $z=5$) the values are $\Psi_c$ and $H_c=H(\Psi_c)$;
in the ${\bf B}_p=0$
region $\Psi_\infty$ and $H_\infty$; and on the
equatorial plane $\Psi_{eq}$ and $H(\Psi_{eq})=0$.
   The numerical values are
 $\Psi_c=0.0384$, $H_c=0.6180$,
$\Psi_\infty =0.133$, $H_\infty=0.637$, and $\Psi_{eq}=0.318$.
  This solution is {\it not} unique.
  For example, we have found an
analogous solution for $k_H \sim 1$.

   The total power output into both the upper and lower
halfspaces is $\dot{E}_{\rm tot} =
B_0^2 r_L^2 c \int_0^{\Psi_{eq}} d\Psi [-H(\Psi)]
=0.152\dot{E}_0$, where $\dot{E}_0\equiv B_0^2 r_L^2 c$.
 In contrast, the total power output of the quasi-spherical
wind solution is $\dot{E}_{tot} \approx B_0^2r_L^2 c$
(e.g., McKinney 2006).

   The jet flow consists of a region of
collimated flux within the light-cylinder $r_L$
where $\Psi \leq \Psi_c=0.0384$,
and a quasi-collimated region outside $r_L$ where
$\Psi_c < \Psi < \Psi_{\infty}=0.133$.
  The power flow in the  collimated jets is
$\dot{E}_{\rm cjet}=
\dot{E}_0 \int_0^{\Psi_c} d\Psi [-H(\Psi)]=
0.0155\dot{E}_0$.
  The power flow in the  quasi-collimated flows is
$\dot{E}_{\rm qcjet}=
\dot{E}_0 \int_{\Psi_c}^{\Psi_{\infty}} d\Psi [-H(\Psi)]=
0.0595\dot{E}_0$.
   The power flow in the disk wind is
$\dot{E}_{\rm dwind}=0.0767\dot{E}_0$.
  Thus, about $10\%$ of the total power goes into
the collimated jet, $40\%$ into the quasi-collimated
flow, and $50\%$ into  the disk-wind.

\subsubsection{Radial Force Balance of Jet}

     For conditions where a collimated
jet exists, the $z-$derivatives in
equation (1) vanish.  The  pulsar
equation can then be written as
\begin{equation}
{d \over dr}B_z^2 +{1\over r^2}{d \over dr}
\bigg[r^2\big(B_\phi^2 -E_r^2\big)\bigg] =0~,
\end{equation}
which expresses the radial force balance.
Multiplying this equation by $r^2 dr$ and
integrating from the axis to $r$ gives
the radial virial equation
\begin{equation}
H^2 =(r^2-1)\left({d\Psi \over dr}\right)^2 +
2\int_0^r {dr  \over r} \left({d\Psi \over dr}\right)^2~,
\end{equation}
following Lovelace, Berk, and Contopoulos (1991).

    Figure 4 shows the two sides of equation
(9) obtained from our numerical solution at $z=5$.
There is a $\sim 10\%$ difference of the two sides of
the equation.
    Calculations in a much larger region are
required to determine whether or not the quasi-collimated
flux outside the light-cylinder becomes
collimated at very large distances.

\subsubsection{Vertical Force Balance of Disk}

   The field solution shown in Figure 3 satisfies
equation (1) accurately everywhere {\it except}
close to the equatorial plane at $r^2 \gg 1$.
The reason for the discrepancy in this region
can be understood by considering the vertical
force balance for $z^2 \ll r^2$ and $r^2 \gg1$ where
$B_z^2 \ll B_r^2$.
   For these conditions equation (1) is
approximately
$$
-r^2{\partial^2 \Psi \over \partial z^2} =
-{1\over 2}{d H^2 \over d\Psi} =
-{1\over 2}{\partial H^2 \over
\partial z} {1 \over \partial \Psi/\partial z}~.
$$
This can  be rewritten as
\begin{equation}
-~{\partial \over \partial \xi}
\left[ -~{1\over 2}
\left({\partial \Psi \over \partial \xi}\right)^2
+~{1\over 2} H^2\right]=0~,
\end{equation}
where $\xi \equiv z/r$.  This expresses the
vertical force balance near the equatorial
plane:  The term $-(\partial \Psi/\partial \xi)^2$
represents the negative pressure of the electric field
and it gives an upward force.
The $H^2$ term is magnetic pressure due to
the toroidal magnetic field and it exerts
a  downward force.

  Figure 5 shows the vertical
profiles of $\Psi$ and $H$.
  These profiles do not satisfy equation (10).
The reason is that equation (1)
omits the plasma kinetic  energy
density in all space including  $z=0$
where the magnetic field reverses direction.
  Therefore we include the influence of
the kinetic energy density
as a term $K(\Psi)/2$ within
the square brackets of equation (10).
  The origin of the kinetic energy is from
the annihilation of the magnetic field.
  [A term of this form can be derived from the
relativistic Grad-Shafranov (GS) equation
of Lovelace et al. (1986).
    The right-hand side of equation (86) of
this work includes a term $-4\pi \rho r^2 dJ(\Psi)/d\Psi$,
where $J=\gamma(1-rv_\phi/c)$ is the Bernoulli
constant and $\gamma$ is the Lorentz factor.
We may write $\rho r^2 =f(\Psi)$ and
$dK/d\Psi =8\pi f(\Psi) dJ(\Psi)/d\Psi$.  The
radial flow speed is
$v_r=-cH^{-1}(\partial \Psi/\partial \xi)$
so that $\rho r^2 v_r ={\rm fct}(\Psi)$. ]
Thus we obtain
\begin{equation}
\left({\partial \Psi \over \partial \xi}\right)^2
=K(\xi)+H^2(\xi)-H_\infty^2~,
\end{equation}
where $H_\infty$ is  the constant value of $H$ at
large $\xi$.
Clearly it is necessary to have $K+H^2 \geq H_\infty^2$.
  Because $H(0)=0$ and
$(\partial \Psi/\partial \xi)_{\xi=0}=0$, $K(0)=H_\infty^2$.
Because $(\partial \Psi/\partial \xi)\rightarrow  0$ as
$\xi$ increases, we have $K\rightarrow 0$ as $\xi$.
  A sufficient condition for having ${\bf E}^2 <
{\bf B}^2=B_r^2+B_\phi^2$ is  $K < H_\infty^2$.
To illustrate the behavior we consider the
dependences $H^2=H_\infty^2[1-\exp(-\xi^2/\xi_H^2)]$
and $K=H_\infty^2\exp(-\xi^2/\xi_K^2)$ with
$\delta \equiv\xi_K /\xi_H>1$.

Figure 6 shows the $\xi$ dependences of $\Psi$,
$H$, and $rE_z$ calculated from equation (11)
for an illustrative case which  maintains the
conditions of Figure 5 of
 $\Delta\Psi= \Psi_{eq}-\Psi_\infty=0.185$
and $H_\infty=0.638$.  For this case we have
taken $\delta =2$.  As a result, $\xi_K =0.286$
which corresponds to a half-angle thickness
of the disk of $\approx 16^\circ$.
   Near the equatorial plane the magnetic
field has the form of an Archimedes' spiral,
namely, $dr/d\phi = -r_L(1-\delta^{-2})^{1/2}$.
    With this modification of the disk configuration,
the global field solution involves {\it no}
delta-function current sheets.

\section{Conclusions}

     This work describes a new solution
for the structure of the
magnetosphere of an aligned rotator
described by the force-free pulsar
equation.
  This is obtained by adjusting the
current flow along the poloidal field
lines which remain within the light-cylinder.
  When this current flow is sufficiently large
a collimated jet forms  within the light-cylinder
and a quasi-collimated flow forms outside of it.
  The solution is not unique.
   The jet is collimated by the toroidal magnetic
field.
  At the same time an anti-collimated
disk-wind forms in the
vicinity of the equatorial plane.
   The anti-collimation is due to the toroidal
magnetic field.
    The vertical force balance of the disk-wind
requires the inclusion of a finite kinetic
energy density near the equatorial plane.
   Roughly one-half of the open field lines go
into the jets and the other half to the disk-wind.
   The total current flow in the jets is equal
and opposite to the current flow in the disk.
  Thus there is no current sheet inside or
outside of the light-cylinder.

  Our jet/disk-wind solution
represents an alternative
to the quasi-spherical wind
solutions where a major part of the
current flow is  in a current sheet.
 Such a current sheet
is unstable to magnetic field annihilation.
   Furthermore, the way in which the configuration
is reached is expected to be important.
  Consider a possible simulation experiment
where the configuration is
reached by dynamical evolution from an initially
non-rotating star.
  The presence, distribution,
and density of the initial plasma is important
in that it is essential for the current
flow as the star is spun up
to a final  rate.
   The value of the density
of the background plasma was found
to have an crucial role in determining
the inflation magnetic loops
threading a differentially rotating
disk in relativistic
particle-in-cell simulations
(Lovelace, Gandhi, \& Romanova 2005).
  The transition from one type of equilibrium
to another remains to be investigated.

We thank A. Spitkovsky, M. Sulkanen, and A.
Timokhin for valuable discussions.
This work was supported in part by NASA grant
 NAG 5-13220 and NSF grant AST-0307817.

\begin{figure*}[t]
\includegraphics[width=5in]{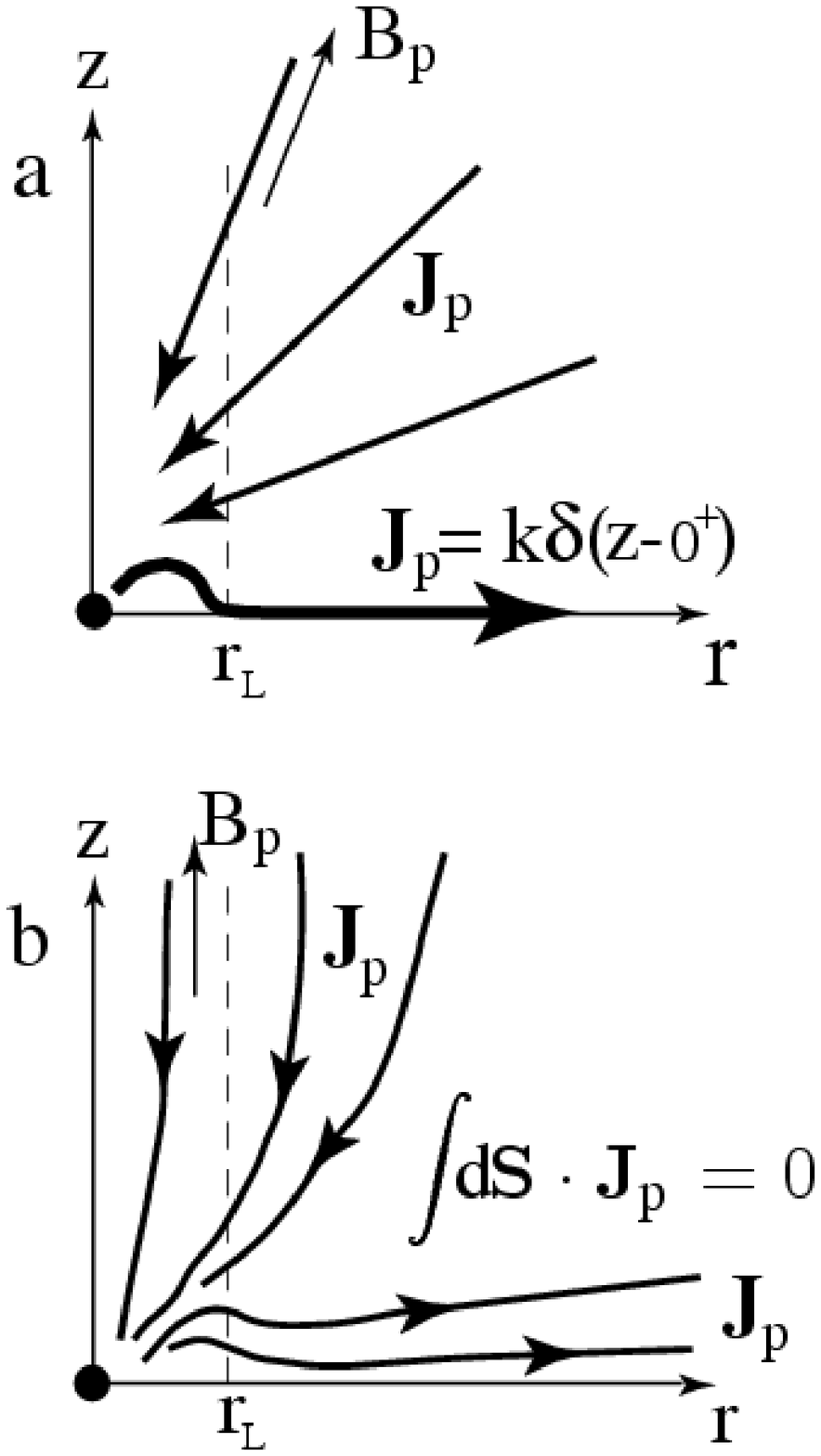}
\caption{({\bf a}) Sketch of
the poloidal current flow in a pulsar
magnetosphere with a quasi-spherical
wind with no jet as proposed by
Contopoulos et al. (1999).
({\bf b})  Sketch of the poloidal current flow
with a collimated jet and a disk-wind.  This
solution is related to that proposed  by
Sulkanen and Lovelace (1990).  }
\end{figure*}

\begin{figure*}[t]
\includegraphics[width=5in]{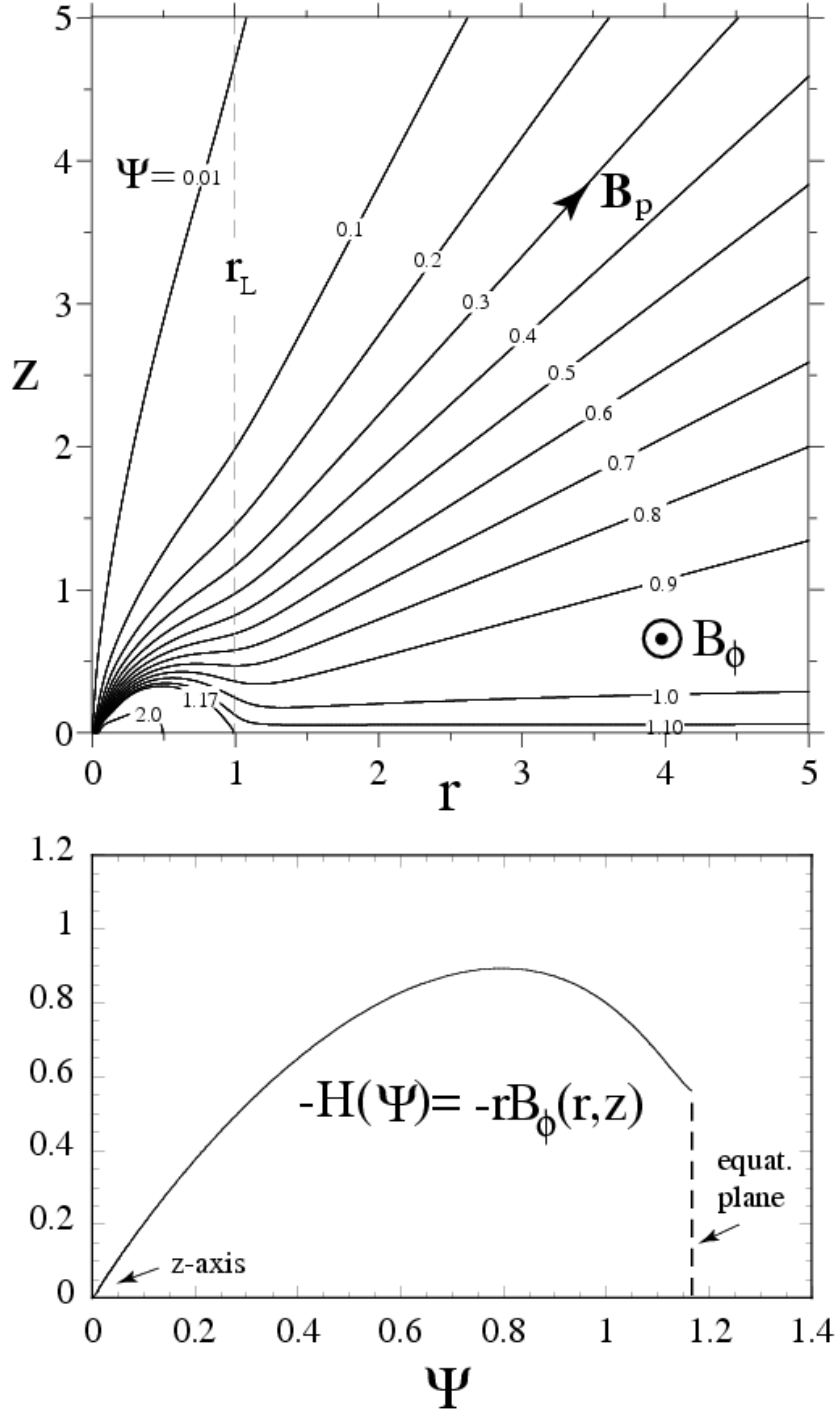}
\caption{Quasi-spherical wind solution
similar to that of Contopoulos et al. (1999).
The jump in $H$ at $z=0$
is related the equatorial poloidal current current
sheet shown in Figure 1a.}
\end{figure*}

\begin{figure*}[t]
\includegraphics[width=5in]{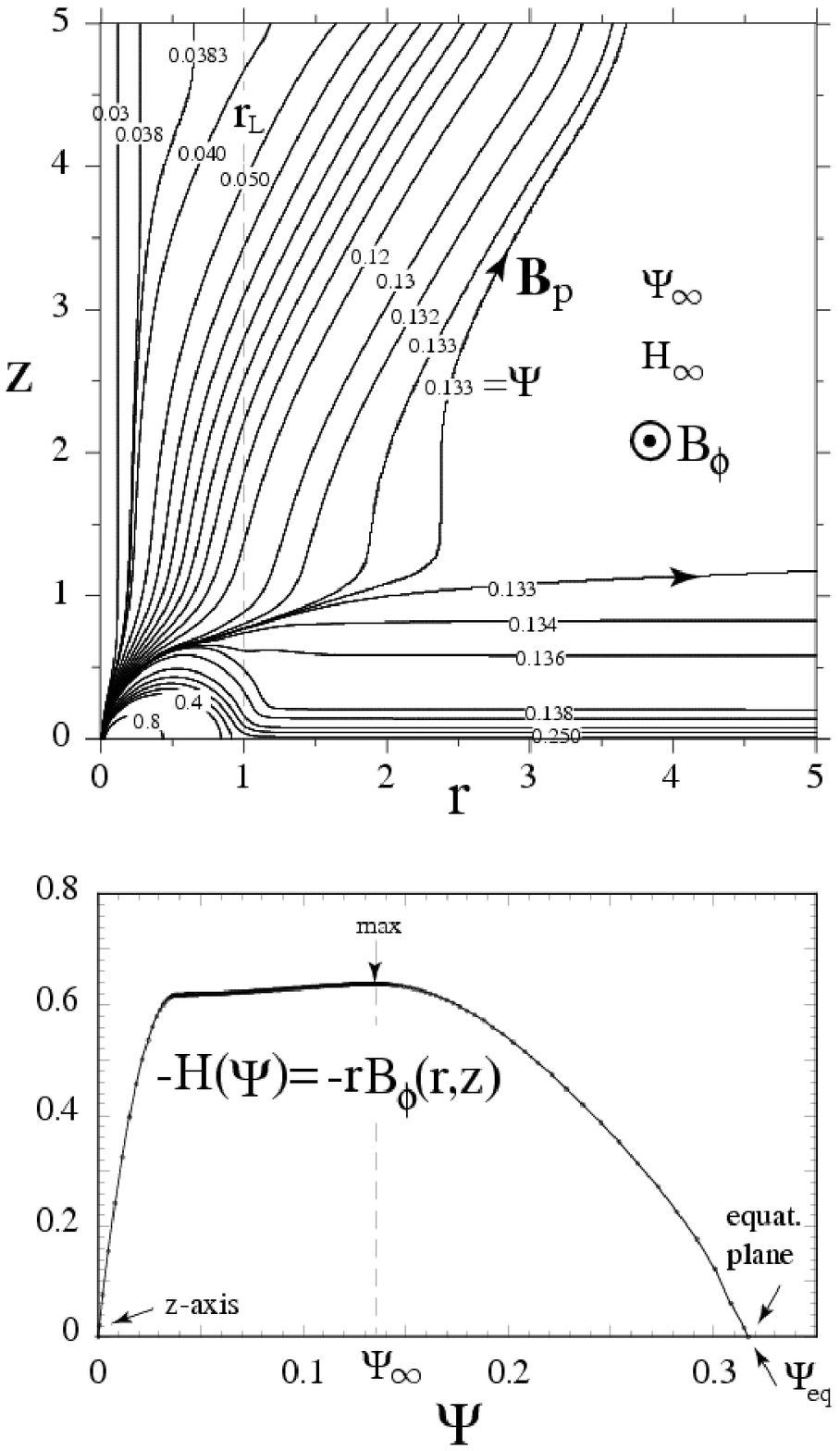}
\caption{Pulsar magnetosphere with
jet obtained by successive iteration
of equation (1) with the constraint
that $\int_0^{\Psi_{eq}} d\Psi {\cal F}=0$.
$H(\Psi)$ within the
light cylinder is given by equation (6) with
$k_H=2.48$ and
 $\beta =0.9995$.  }
\end{figure*}

\begin{figure*}[t]
\includegraphics[width=5in]{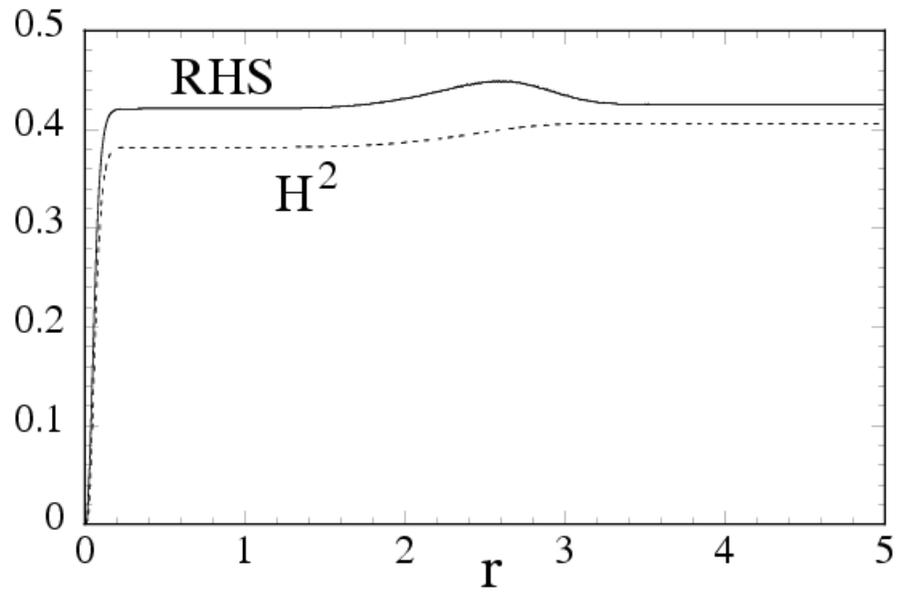}
\caption{Two sides of equation (9) obtained
from our numerical solution at $z=5$.
RHS denotes the right-hand-side of the equation.
}
\end{figure*}

\begin{figure*}[t]
\includegraphics[width=5in]{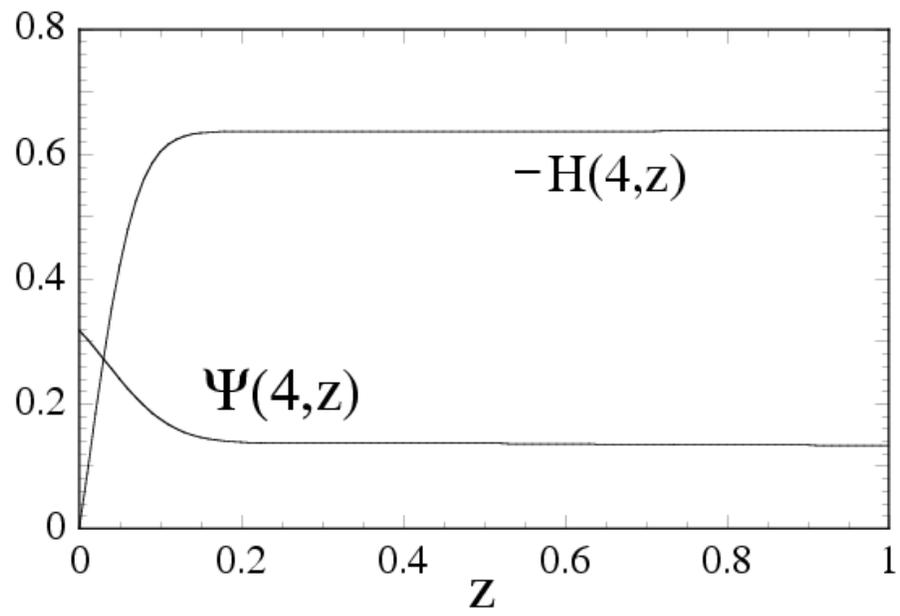}
\caption{Vertical profiles of $\Psi$
and $H$ at $r=4$.  At large $z$,
$\Psi_\infty = 0.133$ and $H_\infty=0.638$.
}
\end{figure*}

\begin{figure*}[t]
\includegraphics[width=5in]{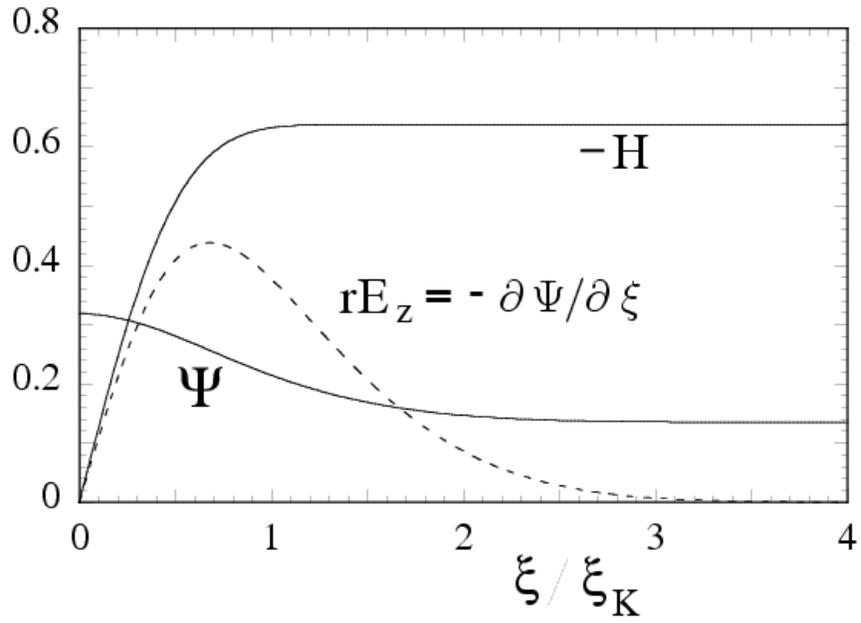}
\caption{Vertical profiles of $\Psi$,
$H$, and $rE_z$ calculated from equation (11)
for illustrative conditions but maintaining
 $\Delta\Psi= \Psi_{eq}-\Psi_\infty=0.185$
and $H_\infty=0.638$. }
\end{figure*}

\end{document}